\documentclass[prb,twocolumn,showpacs,amssymb,amsmath,floatfix]{revtex4}
\usepackage{epsfig}

\usepackage{graphicx}

         \newcommand{\crg}{CeRu$_{2}$Ge$_2$}
         
         \newcommand{\ccs}{CeCu$_{2}$Si$_2$}
         
         \newcommand{\cps}{CePd$_{2}$Si$_2$}
         \newcommand{\cpg}{CePd$_{2}$Ge$_2$}
         \newcommand{\cca}{CeCu$_{5}$Au}

\begin{document}

\title{
Theory of the thermoelectricity of intermetallic compounds with Ce or Yb ions}

\author{
V. Zlati\'c$^1$ and R. Monnier$^2$}

\vskip 2ex

\address{$^1$Institute of Physics, Bijeni\v{c}ka cesta 46, P. O. Box
304, 10001 Zagreb, Croatia}
\address{$^2$ETH H\"onggerberg, Laboratorium f\"ur Festk\"orperphysik,
8093 Z\"urich, Switzerland}


\begin{abstract}

{

\vskip 2ex

The thermoelectric properties of intermetallic compounds with Ce or Yb ions
are explained by the single-impurity Anderson model which takes into 
account the
crystal-field splitting of the 4{\it f} ground-state multiplet, and assumes a
strong Coulomb repulsion which restricts the number of {\it f\/} electrons or
{\it f\/} holes to $n_f\leq 1$ for Ce and  $n_f^{hole}\leq 1$ for Yb ions.
Using the non-crossing approximation and imposing the charge 
neutrality constraint on the local
scattering problem at each temperature and pressure,  the excitation 
spectrum and
the transport coefficients of the model are obtained.
The thermopower calculated in such a way exhibits all the 
characteristic features
observed in Ce and Yb intermetallics.  Calculating the effect of 
pressure on various
characteristic energy scales of the model, we obtain the $(T,p)$ 
phase diagram which
agrees with the experimental data on CeRu$_{2}$Si$_2$, CeCu$_{2}$Si$_2$,
CePd$_{2}$Si$_2$, and similar compounds.
The evolution of the thermopower and the electrical resistance as a 
function of  temperature, pressure
or doping is explained in terms of the crossovers between various 
fixed points of the model
and the redistribution of the single-particle spectral weight within 
the Fermi window.

}
\end{abstract}


\maketitle

\section{Introduction               \label{Introduction} }

\begin{figure}
\center{\includegraphics[width=1 \columnwidth,clip]{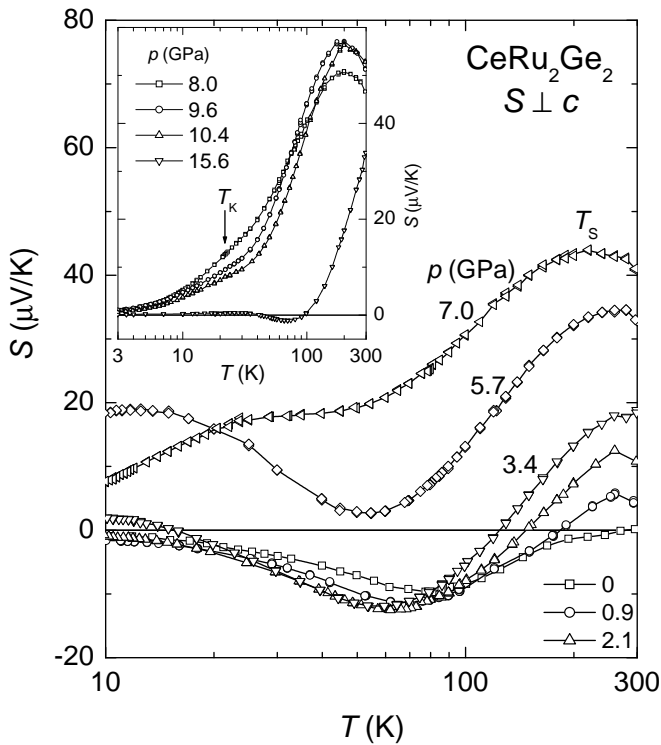}}        
\caption{Temperature dependence of the thermoelectric power $S(T)$ of
       \crg \ for various pressures. $T_K$ and $T_S$ label the centre of
       broad, pressure-induced maxima, related to the Kondo effect and the
       crystalline electric field, respectively. The inset shows $S(T)$
       data of \crg \ in the non-magnetic phase.}
                                                            \label{fig:tepall}      
\end{figure}

The thermoelectric power, $S(T)$, of intermetallic compounds with
Cerium and Ytterbium ions exhibits some characteristic features which
allow the classification of these compounds.
into several distinct groups.\cite{sakurai.96,jaccard.96,zlatic.03}
In the case of Cerium ions, the thermopower of the compounds
belonging to the first group (type (a) systems)  has a  deep negative
minimum at low temperatures
\cite{sakurai.96,jaccard.96,sparn.85,jaccard.85,huo.02}
and a high-temperature maximum, typically between 100 K
and 300 K. At the maximum, $S(T)$ could be either positive or negative,
as shown in Fig. \ref{fig:tepall}. At very low temperatures,
the type (a) systems order magnetically or become superconducting.
The compounds of the second group (type (b) systems)  have a negative
low-temperature
minimum and a positive high-temperature maximum but, in addition, the
thermopower shows a smaller positive peak at lowest
temperatures.\cite{aken.74,cibin.92,sakurai.96,huo.00, sakurai.00}
This second peak is sometimes concealed by a low-temperature phase transition;
for example, in CeCu$_2$Si$_2$ it  becomes visible only in an
external magnetic field which suppresses the superconducting
transition,\cite{sparn.85} and in \crg  \  it shows up when the external
pressure suppresses the magnetic transition.\cite{Wilhelm04}
The experimental evidence is now accumulating that the initial slope
of the thermopower $S(T)/T$  is positive
for this class of (heavy fermion) materials, provided the
measurements are performed at low enough
temperature and with sufficient 
accuracy.\cite{huo.00,sakurai.00,huo.02,behnia.04}
In the third group (type (c) systems), the low-temperature peak is
well pronounced and shifted towards the high-temperature peak.
The main difference with respect to the type (b) systems is that
the sign-change of $S(T)$ does not 
occur.\cite{steglich.77,onuki.87a,sakurai.95,amato.89}
Finally, in some cases (type (d) systems) the thermopower grows
monotonically towards the
high-temperature maximum, and the low-temperature structure appears
only as a shoulder on a broad peak,  or is not resolved at all.
\cite{sakurai.85a,onuki.87a,bauer.91,sakurai.00}

The clue to these various types of behavior  comes from the
high-pressure\cite{Jaccard85,jaccard.88,Wilhelm04,jaccard.92,jaccard.96}
and doping  studies,
\cite{sakurai.96,onuki.87a,sakurai.95,bando.93,gratz.88,sakurai.88,huo.99,ocko.01b}
which show that the thermopower of Cerium compounds
changes continuously from type (a) to type (d). A typical example is
provided by the $S(T)$  of \crg, which is plotted in Fig.\ref{fig:tepall}
as a function of temperature, for various pressures.\cite{Wilhelm04}
At ambient pressure, \crg \ is a type (a) system  with a magnetic ground
state and negative thermopower below 300 K.
An increase of pressure leads to a thermopower with a small positive
peak at low temperatures and an enhanced peak at high temperatures.
A further increase of pressure enhances both peaks, shifts the
low-temperature peak towards the high-temperature one,
and makes the thermopower at intermediate temperatures less negative.
For large enough pressure, the sign-change does not occur at all and for very
high pressure the low-temperature  peak merges with the high-temperature one,
and transforms into a shoulder (see inset to Fig. \ref{fig:tepall}).
The  high-temperature peak grows continuously but its position remains
more or less constant, as $S(T)$ changes from type (a) to (c).
Eventually,  for pressures above 10 GPa, the $S(T)$ assumes the (d) shape.
Here, the initial slope of $S(T)$ decreases and  the position of the maximum
shifts to higher temperatures, but its magnitude does not change as 
pressure increases.
Similar behavior is also seen in the high-pressure  data of,
CeCu$_{2}$Si$_2$,\cite{Jaccard85}
CeCu$_2$Ge$_2$,\cite{jaccard.92,Link96b}
or CePd$_2$Si$_2$.\cite{Link96a}
As regards doping, the substitutions which reduce the volume and make
Ce ions less magnetic, transform $S(T)$ from type (a) to type 
(b),\cite{bando.93}
from (b) to (c),\cite{onuki.87a,sakurai.95} or from (a) 
to(c),\cite{gratz.88,sakurai.88,huo.99}
while the substitutions which expand the volume and make the Ce more magnetic,
transform the thermopower from, say, type (d) to type (c) or from type
(c) to type (b).\cite{ocko.01b}
This variation of shape is an indication that the local environment plays an
important role in determining the magnetic character of  Ce and Yb ions.
  Even at high temperatures, where each 4{\it f} ion is an independent 
scatterer,
  the thermopower of a sample with a high concentration of 4{\it f} 
ions cannot be
  obtained by rescaling the low-concentration data.

The Ytterbium intermetallics can be classified
using the mirror-image analogy with Cerium systems.
This holds because the Yb ions fluctuate between 4{\it f}$^{13}$ and 
4{\it f}$^{14}$,
while  the Ce ions fluctuate between  4{\it f}$^{1}$ and  4{\it 
f}$^{0}$ configurations,
and the dynamics of a single {\it f} hole and a single {\it f\/} 
electron is the same.
A well-defined local moment  leads in  Yb systems to the type (a) behavior,
such that the  thermopower has a negative minimum at high temperatures and a
positive maximum at low temperatures;\cite{alami-yadri.99,nakamoto.99}
the size of the minimum is about the same as the size of the maximum.
The thermopower of (b)-type Yb
systems\cite{casanova.90,trovarelli.99,andreica.99}
mirrors the (b)-type Ce systems. Here, one finds two negative minima
separated by a small positive maximum.
The  type (c)  Yb systems have a nonmonotonic thermopower with a
large (negative)
minimum at high temperatures and a smaller one at low temperatures,
but there is no
sign-change.\cite{casanova.90,andreica.99,alami-yadri.99,trovarelli.99}
Finally, the thermopower with a single negative peak centered around 100 K
\cite{vandaal.74,casanova.90,bauer.91,alami-yadri.99}  mirrors the type (d)
behavior  of Ce systems.
The reduction of volume by pressure or doping\cite{andreica.99,alami-yadri.99}
  stabilizes the magnetic $4{\it f}^{13}$
configuration of Yb ions, and transforms $S(T)$ from, say, type (b) 
to type (a),
from (c) to (b), or from (c) to (a).

The experimental results show that 4{\it f} systems with similar thermopowers
exhibit similarities in other thermodynamic\cite{aviani.01}  and
transport\cite{wilhelm.02,wilhelm.99,demuer.02} properties,
and there is an obvious correlation between
the shape of $S(T)$ and the magnetic character of the 4{\it f} ions.
The thermopower measurements provide a simple and sensitive
tool for characterizing the magnetic state of a 4{\it f} ion in a 
given metallic matrix:
the shape of $S(T)$ changes from the (a)-type in the case of magnetic
Ce (Yb) ions with stable f$^1$ (4{\it f}$^{13}$) configuration to the 
(d)-type for
non-magnetic Ce (Yb) ions which fluctuate between the 4{\it f}$^1$ 
(4{\it f}$^{13}$)
and 4{\it f}$^0$ (4{\it f}$^{14}$) configurations.

We explain the thermoelectric properties of Ce and Yb ions in terms 
of a single-impurity
Anderson model which takes into account the splitting of the 4{\it f} 
states by the crystalline
electric field (CF), and assumes an infinitely large {\it f--f} 
Coulomb repulsion,
which restricts the number of {\it f\/} electrons or {\it f\/} holes 
to $n_f\leq 1$ for Ce
and $n_f^{hole}\leq 1$ for Yb.
We assume that pressure changes the coupling and the relative
occupation of the {\it f} and conduction states, and impose the
charge-neutrality constraint  on the local scattering problem at each
temperature and pressure.
The total charge conservation provides a minimal self-consistency condition
for a poor-man's treatment of pressure effects in stoichiometric compounds.
The excitation spectrum of such a model in the vicinity of various
fixed points, the crossovers induced by
temperature and pressure, and the corresponding effects  on $S(T)$,
the number of {\it f} particles, $n_f(T)$, and the electrical 
resistance, $\rho(T)$,
are calculated by the non-crossing approximation (NCA).
The description of the stoichiometric compounds in terms of an impurity model
is certainly inadequate at low temperatures where the {\it f} electrons become
coherent. The errors due to such an approximation and the low-temperature
errors inherent in the NCA calculations are discussed in detail at the end
of Sec. \ref{Theoretical description}.

Our paper extends the long-standing theory of Coqblin, 
\cite{coqblin.72,coqblin.76}
which described the high-temperature properties of Ce and Yb 
intermetallics  by the
Coqblin-Schrieffer (CS) model with CF splitting,  and its more recent 
version\cite{zlatic.03}
which improved the low-temperature calculations  by rescaling the 
coupling constants.
These previous theories explained the main features of the 
temperature dependence
of the thermopower\cite{coqblin.76,zlatic.03}  and the magnetic 
susceptibility\cite{aviani.01}
but could not describe the pressure effects, because the CS model 
neglects charge fluctuations.
Furthermore, the approximations used to solve the effective high- and 
low-temperature models
cease to be valid at temperatures at which $S(T)$ changes 
sign,\cite{zlatic.03} such that
the shape of $S(T)$ between the two maxima (minima) in Ce (Yb) systems
could only be inferred from an interpolation.

Here,  we consider both the local {\it spin} and {\it charge} 
fluctuations, and provide
the full description of the impurity problem at various pressures and 
temperatures,
from above the CF temperature to below the Kondo temperature,
including the intermediate regime where $S(T)$ changes sign.
Our results explain the shapes (a) to (d) of the thermopower,
which are found in the systems like
CeAl$_3$,\cite{aken.74}  \crg, \cite{Wilhelm04,Wilhelm04a}
CeCu$_{2}$Si$_2$,\cite{Jaccard85}
CeCu$_2$Ge$_2$,\cite{jaccard.92,Link96b} or CePd$_2$Si$_2$, \cite{Link96a}
and the 'mirror-image' shapes found in systems like YbNiSn, YnInAu$_2$, YbSi
or YbCu$_2$Si$_2$.\cite{alami-yadri.99}  We also explain the pressure 
data like on
  \crg \cite{Wilhelm04} or CeCu$_{2}$Si$_2$,\cite{Jaccard85} and the 
chemical pressure
data on Ce$_x$La$_{1-x}$PdSn \cite{huo.99} and 
Ce$_x$La$_{1-x}$Ru$_{2}$Si$_2$\cite{amato.89}
or YbCu$_2$Si$_2$.\cite{andreica.99}

The paper is organized as follows. In Sec. \ref{Theoretical description}
we introduce the model, discuss its limitations, and describe the 
method of solution.
In Sec. \ref{results} we provide the results for the transport 
coefficients of Ce- and Yb-based
intermetallics. In Sec. \ref{Discussion} we discuss
the effects of temperature and pressure on the spectral function,
analyze the fixed-point behavior, and relate the shapes of the thermopower
to the properties of  elementary excitations.
Sec. \ref{Conclusions}, gives the summary and the conclusions.

\section{Theoretical description
\label{Theoretical description}  }

We model the intermetallic compounds by taking as many Ce or Yb
ions per unit cell as required by the structure,
but assuming that the scattering of conduction electrons on a given 
4{\it f} ion does
not depend on other 4{\it f} ions, except through the modification of the
chemical potential.
In other words, we consider an effective impurity model which treats
the 4{\it f} states as
scattering resonances rather than Bloch states but take into account the charge
transfer due to the local scattering and adjust the chemical potential, $\mu$,
so as to maintain the overall charge neutrality of the compound. Such
a description
of the lattice problem applies at temperatures at which the mean free path
of the conduction electrons is short and the scattering is incoherent.
We consider mainly the Ce intermetallics and present only a few preliminary
results for Yb intermetallics.
The Cerium ions are allowed to fluctuate between
the  4{\it f}$^0$ and 4{\it f}$^1$ configurations
by exchanging electrons with the conduction band; the (average)
energy difference between the two  configurations is $|E_f|$ and
the hopping is characterized by the matrix element $V$.
The 4{\it f}$^2$ configuration is excluded, i.e., an infinitely strong
Coulomb repulsion $U$ between {\it f\/} electrons is assumed.
The 4{\it f}$^1$ configuration is represented by $N$ crystal field levels:
there are  ${N}-1$ excited states separated from the CF ground
level by energies $\Delta_i\ll |E_f|$, where $i=1,\ldots,{ N} -1$. The local
symmetry is taken into account by specifying the respective degeneracies
of these levels, ${\cal N}_i$. Thus, the low-energy excitations of Ce 
intermetallics
are modeled by an effective single-impurity Anderson Hamiltonian,\cite{BCW87}
\begin{equation}
                               \label{Hamiltonian}
H_{A}=H_{band}+H_{imp}+H_{mix},
\end{equation}
where $H_{band}$ describes the conduction band,  $H_{imp}$ describes
the CF states, and $H_{mix}$ describes the transfer of electrons 
between 4{\it f} and
conduction states.
In the absence of mixing, the conduction band is described by a 
semielliptical density
of states, $N(\epsilon)$, centered at $E_c^0$ and of half-width $W$, and
the unrenormalized {\it f} states are represented by a set of delta functions
at $E_f^{0}$ and $E_f^{i}=E_f^{0}+\Delta_i$.
The conduction states and the $f$ states have a common chemical 
potential, which is
taken as the origin of the energy axis. The properties of the model
depend in an essential way on the CF splittings and on the coupling constant
$g=\Gamma/\pi |E_f|$,  where $\Gamma=\pi V^2 N(E_{c}^{0})$
measures the coupling strength between the {\it f\/} electrons and 
the conduction band,
$E_f=\sum_{i=0}^{{ N} -1} {\cal N}_i E_f^i/\cal N$,
and $ {\cal N}=\sum_{i=0}^{{ N} -1} {\cal N}_i $ is the total degeneracy.
We assume $E_c^0 > 0$, $E_f < 0$, and $\Gamma, \Delta_i \ll 
|E_f|\simeq W$, i.e., $g\ll 1$.
Since a single {\it f} hole is dynamically equivalent to a single 
{\it f} electron, we
obtain  the results for Yb ions by performing the model calculations
for a 8-fold degenerate {\it f} hole subject to the appropriate CF.

The $g\ll 1$ limit of the Anderson model is controlled by several
fixed points which are well understood.\cite{BCW87}
In the case of a Ce ion with two CF levels split by $\Delta$,
the fixed point analysis can be summarized as follows.
At small coupling, such that  $\Gamma < \Delta\ll |E_f| $,
we find $n_f(T)\simeq 1$ and the model exhibits the Kondo effect.
That is, all the physical properties depend only on the Kondo
temperature, $T_0$, which is
uniquely  determined by $g$, $\Delta$, and the degeneracies of the CF states
(for the NCA definition of $T_0$ see Ref. ~\cite{BCW87} and 
Sec.\ref{Discussion} below).
The low-temperature behavior is characterized by the Fermi liquid
(FL) fixed point, which describes a singlet formed by an
antiferromagnetically coupled {\it f} electron and conduction electron.
An increase of temperature breaks the singlet and gives rise, at about $T_0$,
to a transition to the local moment (LM) fixed point,  which describes
a CF-split {\it f} state weakly coupled to the conduction band.
The effective degeneracy of this {\it f} state is defined
by the lowest CF level.
For $T>\Delta$,  there is a further crossover to another LM fixed
point, which describes the scattering of conduction electrons on a
fully degenerate local moment.
At higher coupling, such that $\Gamma \simeq 2\Delta\ll |E_f| $, the
{\it f} charge is reduced
to $ 0.8 < n_f(T) <0.95$, and the impurity still behaves as a Kondo ion,
but the Kondo scale is much higher than in the case $\Gamma <\Delta\ll |E_f| $.
The two LM regimes are now close together and the crossover from the
FL to the LM regime
occurs at temperatures which are comparable to $\Delta$.
At very high coupling, such that  $2\Delta \leq \Gamma \ll |E_f| $ and
$n_f(T) \leq 0.8$,
the {\it f} ions appear to be non-magnetic at all accessible temperatures
due to the mixing
of the 4{\it f}$^0$ and 4{\it f}$^1$ configurations. In this valence 
fluctuating
(VF) regime, the behavior is non-universal and  changes slightly,
when the calculations are performed for different sets of parameters.
Away from $n_f\simeq 1$, more than one energy scale is needed to
fully characterize the model.
Other CF schemes pertinent to Ce and Yb ions in a different environment
are characterized by similar fixed points.

Our calculations show that the functional form of the response functions
changes at the crossover and that the $g\ll 1$ limit  of the Anderson model
captures all the main features of the experimental results on Ce and 
Yb intermetallics.
To explain the  pressure effects, which changes the thermopower of Ce and Yb
systems in opposite ways,  we  assume that the exchange coupling $g$ increases
in Ce and diminishes in Yb compounds, as pressure increases. This 
difference arises because
Ce fluctuates between 4{\it f}$^{0}$  and 4{\it f}$^{1}$, while Yb fluctuates
between 4{\it f}$^{14}$ and 4{\it f}$^{13}$ configurations, so that 
the pressure-induced
reduction of the number of electrons in the {\it f\/}-shell makes Ce 
ions less magnetic
and Yb ions more magnetic.

In Ce intermetallics, there is a substantial overlap between the
{\it f} wave functions of Ce and those of the neighboring atoms, and
we associate the pressure-induced increase of $g\simeq \Gamma/|E_f|$
with an enhancement of the  hybridization $\Gamma$.
This  enhances the Kondo temperature, and pushes
the system from the Kondo to the VF limit.
In stoichiometric compounds, the pressure-induced reduction of $n_f$ 
is accompanied
by the increase of $n_c$, because the total charge of a given compound,
$n_{tot}=n_c+n_f$, is constant.
The conservation of particles is enforced by adjusting  $\mu$, and since
all the energies are measured  with respect to $\mu$,  this amounts to
shifting $E_f$ and $E_c$ by some amount $\delta\mu(T,\Gamma(p))$.
Thus, we describe the pressure effects for a given Ce compound
by changing $\Gamma$ and keeping $E_c-E_f$ constant.
The changes in the band-width and the CF splitting are neglected.

In Yb intermetallics, the  {\it f} states are more localized than in 
Ce systems,
and we assume that the decrease in the radius of the 4{\it f} shell as it loses
charge at elevated pressure is sufficient to compensate for the 
increase in hybridization
brought about by the reduction in unit cell volume.
The  reduction of $g\simeq \Gamma/|E_f|$ in Yb compounds  is achieved 
through an enhancement
of the hole binding energy $E_f$ as the neighbouring ions get closer 
to the rare earth,
while $\Gamma$  remains essentially  constant.
This reduces the Kondo temperature, and drives the system towards the 
Kondo limit.
Since $\Gamma$  is treated as a material-specific constant, we model the
pressure effects in Yb systems by shifting $E_f$ and solving for  $E_c$,
so as to preserve $n_{tot}^{hole}$.
This procedure shifts $E_f$ and $E_c$ by different amounts and makes the
separation $E_c-E_f$ pressure dependent. However, when temperature is
changed at constant pressure, the charge neutrality is enforced in the same way
as for Ce compounds, by shifting the chemical potential without changing the
separation $E_c-E_f$. The changes in the bandwidth and the CF
splitting are neglected. Describing the pressure effects in such a way,
we can calculate the response functions of the model for any value of
the external parameters, and study the transitions between various 
fixed points.

The electrical resistivity and the thermopower of the single-impurity
Anderson model are
obtained from the usual expressions,\cite{Mahan81}
\begin{equation}
{\rho_{mag}}=\frac{1}{e^2 L_{11}},
\label{eq: conductivity}
\end{equation}
\begin{equation}
S=-\frac{1}{|e|T}\frac{L_{12}}{L_{11}},
                                  \label{eq: thermopower}
\end{equation}
where $ L_{11}$ and $ L_{12}$ are given by the static limits of the
current-current and current-heat current correlation functions,
respectively. In the absence of nonresonant scattering the vertex corrections
vanish and the transport integrals can be written 
as,\cite{BCW87,Mahan98,costi.94}
\begin{equation}
                                           \label{eq: lij_final}
L_{ij}
=
\frac{\sigma_0}{e^2}\int_{-\infty}^{\infty}d\omega
\left (
-\frac{df(\omega)}{d\omega} \right ) \tau(\omega)\omega^{i+j-2},
\end{equation}
where $\sigma_0$ is material-specific constant,
$f(\omega)=1/[1+\exp(\omega/k_B T)]$ is the Fermi function,
$1/\tau(\omega)$ is the conduction-electron scattering rate,\cite{BCW87}
\begin{equation}
                                            \label{eq: tau}
\frac 1 {\tau(\omega)}
=
c {\cal N} \pi V^2 A(\omega),
\end{equation}
$A(\omega)=\mp \frac{1}{\pi}\mbox{Im}\; G_f(\omega \pm i 0^{+})$ is
the {\it f\/}-electron spectral function, $G_f(z)$ is the Green's function,
and $c$ is the concentration of {\it f} ions.
Eqs.(\ref{eq: lij_final})  and (\ref{eq: tau}) show clearly that
the sign and the magnitude
of $S(T)$ are determined by the spectral weight within the Fermi
window  (F window),
i.e., by the shape of $A(\omega)$ for $|\omega|\leq 2k_BT$.
The sign of $S(T)$ is positive if the F window shows more states above than
below the chemical potential, and is negative in the opposite  case.
The difficult part is to find $ G_f(\omega \pm i 0^{+})$ and, here,
we solve this
problem by the NCA, following closely Refs.~\cite{BCW87,Monnier90},
where all the technical details can be found.  The main difference 
with respect to
these NCA calculations is that we take $c=1$ and enforce the overall 
charge neutrality.

A detailed comparison with the experimental data shows that the
transport coefficients
obtained from the single-impurity Anderson model have all the
hallmarks of the experiments,
but discrepancies appear at low temperatures. This indicates the
limitations of our approach which should be considered before
presenting the NCA results.

There are two main causes for the breakdown of the single-impurity
model and the
NCA calculations. First, at temperatures much below $T_0$ the NCA
spectral function
develops an unphysical spike, such that the resistivity and the
thermopower become
artificially enhanced. This error becomes particularly severe at high
pressure, because
the characteristic scale $T_0$ increases very rapidly with $\Gamma$
and the non-analytic
NCA spike appears at rather high temperatures.  The unphysical enhancement of
the low-frequency part of  $A(\omega)$ reduces the integral for $L_{11}$,
which is strongly underestimated at low-temperature. The integral
$L_{12}$ is less affected
by this pathology,
because it has an additional $\omega$ factor which removes the states
within the Fermi window.
Thus, the overall shape of $S(T)\simeq L_{12}^{NCA}/L_{11}^{NCA}$ seems to be
qualitatively correct,  even though the low-temperature part of the
curve has an unphysical
enhancement. These difficulties are well known ~\cite{BCW87} and
relatively easy to
resolve in the Kondo limit,
where the model has a unique low energy scale, $T_0$. We can find $T_0$ in
the LM regime,
where the NCA is reliable, and infer the low-temperature behavior
from the universal power
laws which hold in the FL regime. Combining the NCA results and FL
theory, we can discuss
the experimental data at all temperatures at which the single-ion
approximation holds.

A more serious problem is that, in stoichiometric compounds, the
{\it f} electrons become coherent
at low enough temperatures. This leads to a magnetic transition in
CeRu$_{2}$Ge$_2$,
the formation of a heavy FL in CeRu$_{2}$Si$_2$, superconductivity
in CeCu$_{2}$Si$_2$, or some more complicated ground states.
At high pressure, the coherence sets in at very high temperatures, as
revealed by
low values of the electrical resistance.
The onset of coherence (like the NCA pathology) has its main impact
on the low-energy
states, giving $ L_{11}^{lattice} \gg L_{11}^{impurity}$, so that the
impurity result
badly overestimates the low-temperature electrical resistance.
However, the considerations for the periodic Anderson model\cite{Mahan98},
or other models with on-site correlation\cite{freericks.03},
show that the integral for $L_{12}$ also contains an additional $\omega$ factor
which reduces the contribution of the low-energy coherent states to $L_{12}$,
like in the single impurity case. Thus,  our results for $S(T)$ provide
a qualitative description of the experimental data at temperatures
well below the onset of coherence, but the calculated values of $S(T)$
around $T_0$ are overestimated.

We mention also that the analysis of the doping effects in terms of
''chemical pressure'' is
not complete, because doping might give rise to a charge transfer or
change the character
of the ground state, and that the mirror-image analogy between Ce and
Yb systems holds
for the resonant scattering but may be lost in the presence of any
additional scattering channel.
Despite these drawbacks, the NCA solution of the Anderson model provides
a surprisingly accurate description for a large body of experimental data
above the magnetic or superconducting transition temperature,
and at low to moderate pressure.

\section{Thermoelectric power results
\label{results}  }

In this section we present the NCA results for the transport coefficients,
describe the thermopower and the electrical resistance due to Ce ions 
in some detail,
show the results for the {\it f} occupation, present a few 
preliminary results for the
thermopower of Yb, and compare our results with the experimental data.

Our strategy for Ce intermetallics is to illustrate the behavior of
one particular compound as a function of temperature at various pressures.
The compound is characterized at ambient
pressure and high temperature by an initial parameter set
$\{W, E_c^0,{N}, \Gamma, E_f^0,  \Delta_i, {\cal N}_i\}$,
where $E_c^0>0$ and $E_f^0<0$ are measured with respect to $\mu$ and the
high-temperature limit is defined by temperature $k_BT_\Delta=\Delta_{{N}-1}$.
For given values of $\Gamma$ and $E_c^0-E_f^0$,
we start the calculations at $T=T_{\Delta}$ and find 
$\delta\mu(T_\Delta,\Gamma)$
such that the total charge is the same as the one obtained for $\Gamma=0$.
(In the absence of coupling we have $n_f=1$ and obtain  $n_c$ by integrating
the unperturbed density of states.)
At high temperatures, the {\it f\/}-state is almost decoupled from 
the conduction band,
the renormalization of the parameters is small, and the numerics 
converge very fast.
We then reduce the temperature, find the new shift $\delta\mu(T,\Gamma)$
ensuring the charge conservation, and calculate the response functions
for the resulting values of $E_c$ and $E_f$. This process is 
continued until the NCA
equations break down at $T\ll T_0$.
To model the same system at different pressure, we change $\Gamma$,
find again $\delta\mu(T_\Delta,\Gamma)$, and repeat  the same procedure
as at ambient pressure for $T < T_\Delta$. Note,  $n_{tot}$  and $E_c-E_f$
are conserved at all temperatures and pressures.
\begin{figure}
\center{\includegraphics[width=1\columnwidth,clip]{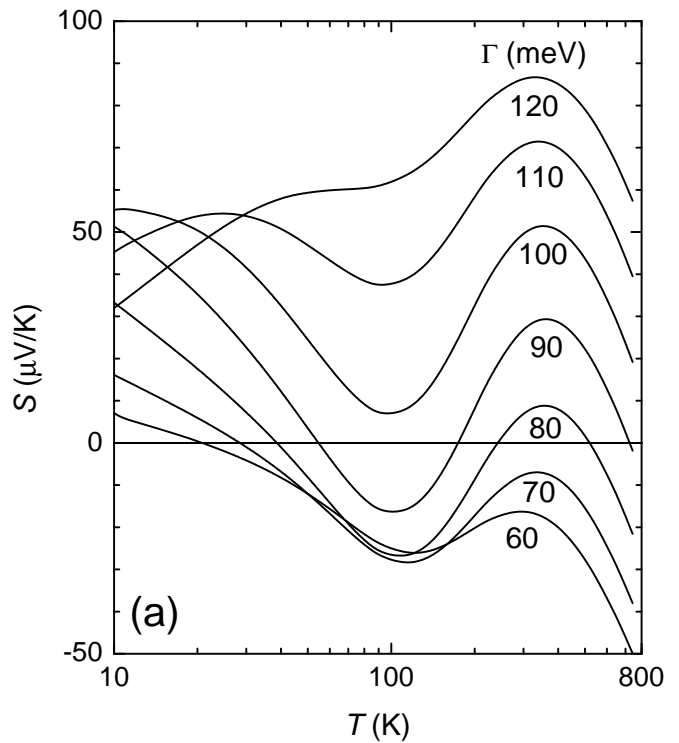}}          
\caption{Thermopower of Ce ions calculated by the NCA for  the CF splitting
$\Delta=0.07$ eV is plotted as a function of temperature for several 
values of the
hybridization strength $\Gamma < 2 \Delta$, as indicated in the figure.
The bottom curve, $\Gamma$=0.06 eV, corresponds to ambient pressure. }
                                           \label{fig:theo_tep_low}   
\end{figure}

\begin{figure}
\center{\includegraphics[width=1\columnwidth,clip]{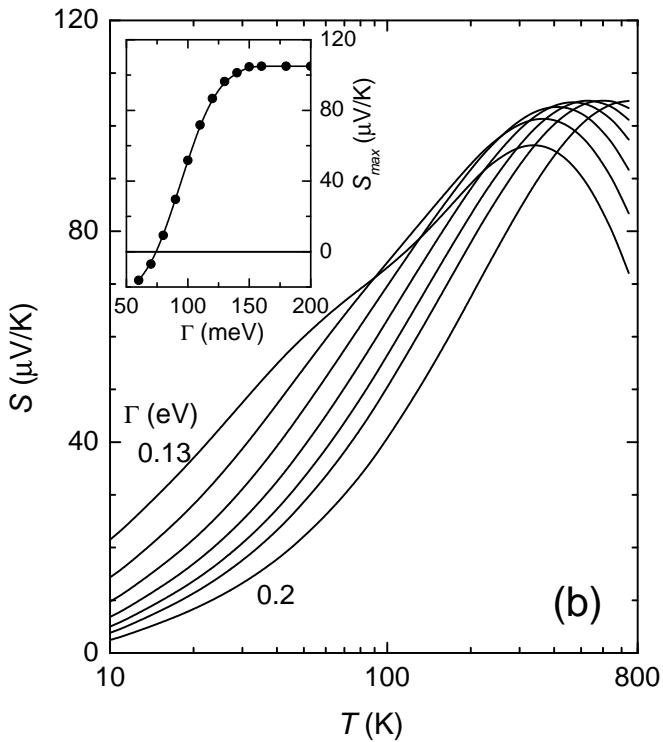}}    
\caption{Thermopower calculated by the NCA for  the CF splitting
$\Delta=0.07$ eV is plotted as a function of temperature for
several large values of the hybridisation strength
$\Gamma$,  increasing from 0.13 eV to 0.2 eV. The bottom curve,
$\Gamma$=0.20 eV, is for the highest pressure.
The inset shows the high-temperature
maximum of $S(T)$, plotted as a function of $\Gamma$. }
                                             \label{fig:theo_tep_large}  
\end{figure}

As a numerical example we consider a semielliptical conduction band
of half-width $W=4$ eV, centered at  $E_c^0=0.7$ eV, and a {\it f} 
state split into
a doublet and a quartet\cite{quartet} by the CF  with $ \Delta=0.07$ eV.
We take $n_{tot}=n_c+n_f$=5.6301 electrons per ion (0.9383 electrons
for each one of $\cal N$ 'effective spin' channels).
The transport coefficients are calculated for the hybridization 
strength changing
from 0.06 eV  to 0.20 eV, i.e., for  $\Gamma$  varying from $\Gamma < \Delta$
to $\Gamma > 2\Delta$.
The single-particle excitation spectra corresponding to these parameters
are discussed in Sec.~\ref{Discussion}.

Using the procedure outlined above we obtain for $S(T)$ the results
shown in Figs.~\ref{fig:theo_tep_low} and ~\ref{fig:theo_tep_large}.
The calculated curves  exhibit all the shapes (a) to (d) found in the
experimentas and give $S(T)$ of the right magnitude, except at low
temperatures  where the calculated peak is too large with respect
to the experimental data.
As discussed already, the reason for this discrepancy is that the NCA
overestimates the Fermi-level scattering rate for $T\ll T_0$, and
that we neglected the coherent scattering, which sets in at temperatures
of the order of $T_0$.
Thus, our low-temperature result for $L_{11}$ is artificially reduced,
which makes $S(T)$ too large. The sign and the topology of the $S(T)$
curves do not seem to be affected by this error.

\begin{figure}
\center{\includegraphics[width=1 \columnwidth,clip] {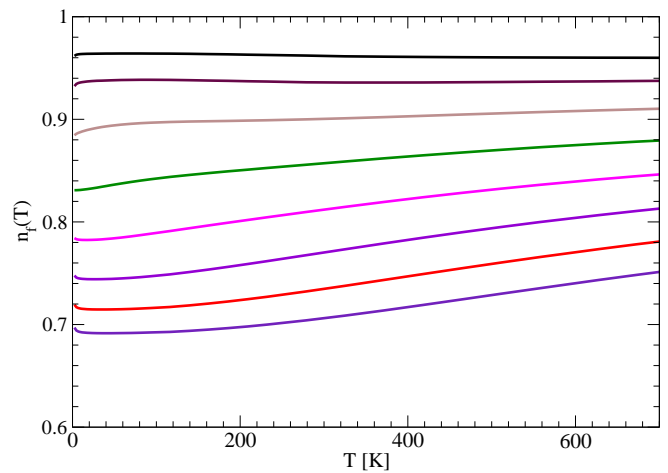}}                
\caption{f-electron number, $n_f$, calculated by the NCA for  the CF splitting
$\Delta=0.07$ eV is plotted as a function of temperature
for several values of the hybridisation strength $\Gamma$.
For the uppermost curve $\Gamma$=0.06 eV and then
it increases in steps of 0.02 eV.
At the bottom curve $\Gamma$=0.20 eV. }
                                  \label{fig:nf}      
\end{figure}

For $\Gamma\leq \Delta$, we have $T_0 < 5$ K and $n_f \geq  0.95$,
and obtain $S(T)$ with two well separated peaks, as shown by the $\Gamma$=60
and $\Gamma$=70 meV curves  in  Fig.~\ref{fig:theo_tep_low}.
The high-temperature peak is centered at $T_S\simeq T_\Delta/2$
and for our choice of parameters  $S_{max}=S(T_S) <0$.
The low-temperature maximum is at  about $T_{0}\ll T_S$ and $S_0=S(T_0) > 0$.
The thermopower between the two maxima is mainly negative.
Since most of the type (a) and (b) systems order magnetically or 
become superconducting
above $T_0$, the low-temperature peak is not shown in 
Fig.~\ref{fig:theo_tep_low}
for $\Gamma\leq 100$ meV.
A small increase of $\Gamma$ (due to, say, an increase of pressure)
reduces $n_f$ (see Fig.~\ref{fig:nf}), enhances  $T_0$ and $S_{max}$,
and expands the temperature range in which $S(T)$ is positive.
Such a behavior, which is typical of Kondo systems with small $T_0$,
is in a qualitative agreement with the thermopower of the type (a) systems
described in Sec.\ref{Introduction}, and  with the data on \crg 
\cite{Wilhelm04}
at low pressures (below 4 GPa) and above the ordering temperature,
as shown in Fig.\ref{fig:tepall}.

For $\Delta < \Gamma < 2\Delta$, we have  $T_0 < $150 K and $n_f \geq $ 0.8,
and still obtain  $S(T)$  with the  two maxima. But  $S_{max}$ is now positive,
the value of $S_0$ is enhanced, the  temperature interval in which 
$S(T) < 0$  is reduced,
and $S(T)$ at the minimum is less negative than for smaller  $\Gamma$.
As we increase $\Gamma$ (by increasing pressure),  the sign-change of $S(T)$
is removed, $S_0$ and $S_{max}$ are further enhanced, but $T_S$ is not changed.
The two peaks are coming closer together and are merging eventually.
These features  are typical of Kondo systems with moderate $T_0$,
say $T_0\geq 10$ K, and are in a qualitative agreement with the data
on the type (b) and (c) systems mentioned in Sec.~\ref{Introduction}.
They are also shown by the \crg \ data\cite{Wilhelm04} at 
intermediate pressures,
(see the curves in Fig.~\ref{fig:tepall} for pressure above 3.4 GPa.)

At $\Gamma \approx 2\Delta$, the system enters the
VF regime, $S(T)$ acquires a single maximum at $T_S$,
with a shoulder on the low-temperature side.
Even this shoulder vanishes, when $\Gamma$ is increased further,
as shown in Fig.~\ref{fig:theo_tep_large}. For $\Gamma > 2\Delta$
the thermopower is of the type (d), with a single peak which
is much steeper on the high- than on the low-temperature side.
This peak shifts to higher temperatures with increasing $\Gamma$ (pressure)
and $S_{max}$ saturates
(see inset to  Fig.~\ref{fig:theo_tep_large}); the initial slope of $S(T)$
decreases continuously with $\Gamma$.
Such a behavior is in a qualitative agreement with the thermopower data on
valence fluctuators\cite{sakurai.85a,onuki.87a,bauer.91}
and with the high-pressure data on \crg \cite{Wilhelm04}
(cf. inset to Fig.~\ref{fig:tepall}) and \ccs\cite{jaccard.85}.
However, a large discrepancy appears between theory and experiment at 
low temperatures,
because $A(\omega)$ is overestimated for $\omega\simeq 0$,
which makes the NCA curves larger than the experimental ones. A possible
correction of the initial $S(T)/T$ values is discussed below.

The {\it f\/}-electron number $n_f$, calculated for the parameters used
in Figs.~\ref{fig:theo_tep_low} and ~\ref{fig:theo_tep_large},
is plotted in Fig.~\ref{fig:nf}  as a function of temperature.
The overall temperature dependence is rather slow, but two different types
of behavior can still  be seen.
For $\Gamma <  2\Delta\ll -E_f$, we find that $n_f $ is nearly independent of
temperature and close to 1, which  is typical of Kondo systems.\cite{n_f=0}
For  $\Gamma >  2\Delta$,  we find that $n_f $ is less than 0.8 and nearly
constant at low temperatures but at about $T\simeq \Delta/3k_B$
(271 K for $\Delta$=0.07 eV) there is an increase followed by the saturation
at high temperatures.
Considered as a function of $\Gamma$ (pressure), $n_f(\Gamma)$
shows different behavior at high and low temperatures.
At high temperatures $n_f$ decreases uniformly as $\Gamma$ increases.
At low temperatures $n_f$ does not change much for  $\Gamma\ll \Delta$ and
$\Gamma\gg \Delta$, but drops rapidly around $\Gamma\simeq 2\Delta$,
indicating the crossover from the Kondo to the VF regime.

\begin{figure}
\center{\includegraphics[width=1 \columnwidth,clip]{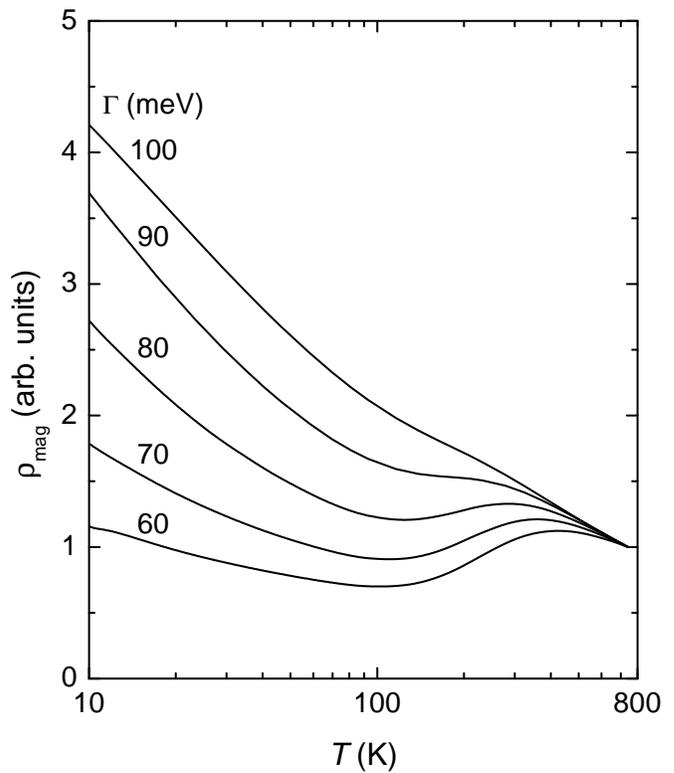}}      
\caption{Electrical resistivity vs. temperature
calculated by the NCA for the CF splitting $\Delta=0.07$ eV
and for several values of the hybridization strength $\Gamma$,
as indicated in the figure.}
                                                       \label{fig:theo_rho}   
\end{figure}

The electrical resistivity, $\rho_{mag}(T)$, obtained for small
and intermediate values of $\Gamma$, is shown in Fig.~\ref{fig:theo_rho}.
The interesting feature is the high-temperature maximum, which appears
for $\Gamma < \Delta$, and correlates very well with the maximum in $S(T)$.
For temperatures below the maximum, $\rho_{mag}(T)$ drops to a minimum
and then rises logarithmically as $T_0$ is approached.  This minimum 
and the subsequent
low-temperature upturn are of a purely electronic origin and
appear in systems with small $T_0$ and large CF splitting.
In these systems, one can follow the evolution of the two peaks in 
$\rho_{mag}(T)$,
and observe the disappearance of the minimum with the application of pressure.
For example, in
\crg,\cite{Wilhelm04} \cpg,\cite{wilhelm.02} \cps,\cite {demuer.02} 
or \cca\cite{wilhelm.99}
the minimum becomes more shallow, transforms into a shoulder,
and vanishes at high enough pressure.
As discussed already, the NCA overestimates the low-temperature scattering and
distorts the relative magnitude of the high- and low-temperature peaks.
In addition, the single-ion approximation always gives
$\rho_{mag}(T)$  which saturates at low temperature and
cannot explain the low-temperature reduction of $\rho_{mag}(T)$,
which is seen in stoichiometric compounds below the onset of coherence.
The electrical resistance of Ce-based Kondo systems at very high pressure,
and the ambient pressure data of valence fluctuators, cannot be described
by the NCA solution. In these systems,
the scattering on {\it f} ions remains coherent up to rather high
temperatures and the NCA solution
is valid only above the high-temperature maximum and is not
physically relevant.

\begin{figure}
\center{\includegraphics[width=1 
\columnwidth,clip]{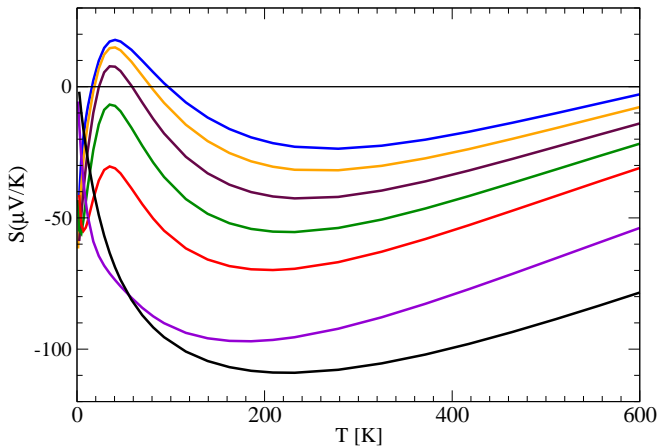}}                  
\caption{Thermopower due to Yb ions, obtained by the NCA for four CF doublets,
is plotted as a function of temperature for various values of $E_f^0$,
i.e. for various pressures. Starting from the uppermost curve, we 
show $S(T)$ for
$-E_f^0$=1.00; 0.95; 0.90; 0.85, 0.80, 0.70, and 0.60  eV, respectively.
The bottom curve ($E_f$=-0.6 eV) corresponds to the lowest pressure. }
                                             \label{Yb_tep_NCA}    
\end{figure}

To illustrate the situation in Yb intermetallics, we perform a 
generic calculation
for a semielliptical (hole) conduction band and four CF doublets.
The NCA equations are solved for a single {\it f} hole (${\cal N}=8$).
Following the procedure outlined in Sec. \ref{Theoretical description},
we start the calculations at ambient pressure and temperature
$T_\Delta$, such that the {\it f} state is almost free, and calculate
$n_{tot}^{hole}=n_c^{hole}+n_f^{hole}$ for the initial parameter set
$\{W, E_c^0,  { N}, \Gamma, E_f^0,  \Delta_i, {\cal N}_i\}$.
At lower temperatures we shift $E_c$ and $E_f$ with respect to $E_c^0$
and $E_f^0$ by $\delta\mu$, so as to conserve  $n_{tot}^{hole}$, and
calculate the response functions for this new parameter set; this procedure
does not change $E_c-E_f$.
At a higher pressure, we start again at $T=T_\Delta$, change $E_f$ so as to
increase $n_f^{hole}$ and find $E_c$ which conserves  $n_{tot}^{hole}$.
Since $\Gamma$ is not  changed by this procedure,  we now have
$E_c-E_f\neq E_c^0-E_f^0$.
For temperatures below $T_\Delta$, the properties of the system are calculated
by the same procedure as at ambient pressure, i.e.,  $n_{tot}^{hole}$ 
is conserved
by shifting $E_c$ and $E_f$ by the same amount $\delta\mu$.

Taking $W=4$ eV, $E_c^0=1.0$ eV, $\Gamma$=0.08 eV, $E_f^0=-0.8$ eV,
and three excited CF doublets at $\Delta_1$=0.02 eV, $\Delta_2$=0.04 eV,
$\Delta_3$=0.08  eV, respectively, we find $n_{tot}=6.444$ at $T=T_\Delta$.
The $S(T)$ obtained for $E_f$ ranging from  $E_f=-0.6$ eV to $E_f=-1.0$ eV,
is shown in Fig. ~\ref{Yb_tep_NCA}. We recall, that an increase
of pressure makes $E_f$ more negative.
For $E_f=-0.6$ eV, the thermopower shows a deep minimum, typical of 
Yb ions in the VF state.
For $E_f=-0.8$ eV,  $S(T)$ develops a small maximum,
which separates the high-temperature minimum at $T_S$
and the low-temperature minimum at $T_0$.
By making $E_f$ more negative we shift  $T_0$ to lower values,
as shown in Fig. \ref{Yb_tep_NCA}  by the curves obtained for
$E_f=-0.8$ eV, $E_f=-0.85$ eV, and $E_f=-0.9$ eV, respectively.
The low-temperature range in which $S(T)$ is negative
shrinks with pressure, in agreement with the experimental 
data.\cite{alami-yadri.99}
However, for $E_f$ much below $\mu$, the NCA calculations break down before
this minimum is reached.  As regards the value of $S(T)$ at the maximum,
it is negative at first but it becomes positive as pressure increases,
i.e.,  the thermopower changes from (c)-type to (b)-type.
At very high pressures, such that the $n_{tot}^{hole}\simeq 1$, the thermopower
is dominated at low temperatures by a large positive peak and  at 
high temperatures
by a negative minimum, which is typical of Yb-based  systems with a 
small Kondo scale.
The shape of  $S(T)$ is directly related to the magnetic character of 
Yb ions and
our calculations explain the qualitative features of the thermopower
of YbAu$_2$, YbAu$_3$, \cite{nakamoto.99} and YbSi, 
YnNi$_2$Si$_2$,\cite{alami-yadri.99}
which are of the (a)-type,
of Yb$_2$Ir$_3$Al$_9$,\cite{trovarelli.99} 
YbAuCu$_4$,\cite{casanova.90} and YbNiSn,\cite{alami-yadri.99}
which are of the (b)-type,
of YbPdCu$_4$,\cite{casanova.90} and Yb$_2$Rh$_3$Al$_9$,\cite{trovarelli.99}
which are of the (c)-type,
and of YbAgCu$_4$,YbPd$_2$Cu$_2$,\cite{casanova.90} and 
YbInAu$_2$\cite{alami-yadri.99}
which are (d)-type.
The pressure effects in YbSi,\cite{alami-yadri.99} and
the chemical pressure effects in YbCu$_2$Si$_2$\cite{andreica.99}
are also in a qualitative agreement with our results.

An estimate of the low-temperature properties of the single-impurity
model can be obtained by combining  the NCA results with the 
universal  FL laws.
The Sommerfeld expansion of transport coefficients in  Eq.(\ref{eq: lij_final})
gives\cite{costi.94}
  \begin{equation}
\lim_{T\to 0} \frac{\Gamma S(T)}{k_B T}
=
\frac{\pi^2 k_B}{ 3|e|} \sin( \frac{2\pi}{ {\cal N}_0} n_{f})  Z,
                              \label{resonant_SIAM}
\end{equation}
where  ${\pi^2 k_B}/{ 3|e|}$=283.5 $\mu$V/K and
$Z$ is  the enhancement factor defined by the Fermi-level
derivative of the {\it f\/}-electron self-energy,
$Z=\left[1-{\partial \Sigma}/{\partial \omega}\right]_{\omega=T=0}$.
$\Gamma/Z$ is related to the Kondo scale by a factor of order 1.
Setting $T_0=\Gamma/Z$,  and
using for $n_f(T=0)$ the NCA results of Fig.~\ref{fig:nf},
we can estimate from Eq.(\ref{resonant_SIAM}) the initial slope of the
curves plotted in Figs.~\ref{fig:theo_tep_low}  and  ~\ref{fig:theo_tep_large}.
The obtained values, which go from 32 $\mu$V/K$^{2}$ for $\Gamma$=60 meV
to 1 $\mu$V/K$^{2}$ for $\Gamma$=130 meV  and 0.1 $\mu$V/K$^{2}$ for 
$\Gamma$=200 meV,
are in the range reported recently for various Ce-based heavy
fermions and valence fluctuators (see Table 1 in ref. ~\cite{behnia.04}).
The initial slope of $S(T)$ decreases as we move from the Kondo to 
the VF limit,
in agreement with pressure experiments. \cite{Wilhelm04}
At higher temperatures, the non-linear corrections reduce $S(T)$ and
give rise to a maximum at $T_0$. These non-linear corrections are 
non-universal, and
a large slope does not necessary translate into large $S(T)$ at $T_0$.
The corresponding calculations for the  Yb curves plotted in Fig. 
~\ref{Yb_tep_NCA}
show that an application of pressure enhances $S(T)/T$.

The initial slope of $S(T)$ can also be related to the $\gamma$ 
coefficient of the
specific heat, with the important result that the ratio
$S(T)/\gamma T= (2\pi/|e|{\cal N}_0) \mbox{cot}(\pi n_{f}/{\cal N}_0)$
is independent of $Z$.
This expression and Eq.(\ref{resonant_SIAM}) are valid in the 
single-impurity regime,
and it is not obvious that they would produce the correct results for 
stoichiometric compounds.
However, the characteristic energy scales of a coherent FL (inferred from the
experimental data on the initial slope of the thermopower, or the 
specific heat coefficient)
do not seem to be much different from the single-ion scale ${T_0}$ of 
the LM regime
(inferred, say, from the peak in the thermopower or the Curie-Weiss 
temperature).

\section{Discussion  of spectral properties
                                    \label{Discussion} }
\begin{figure}
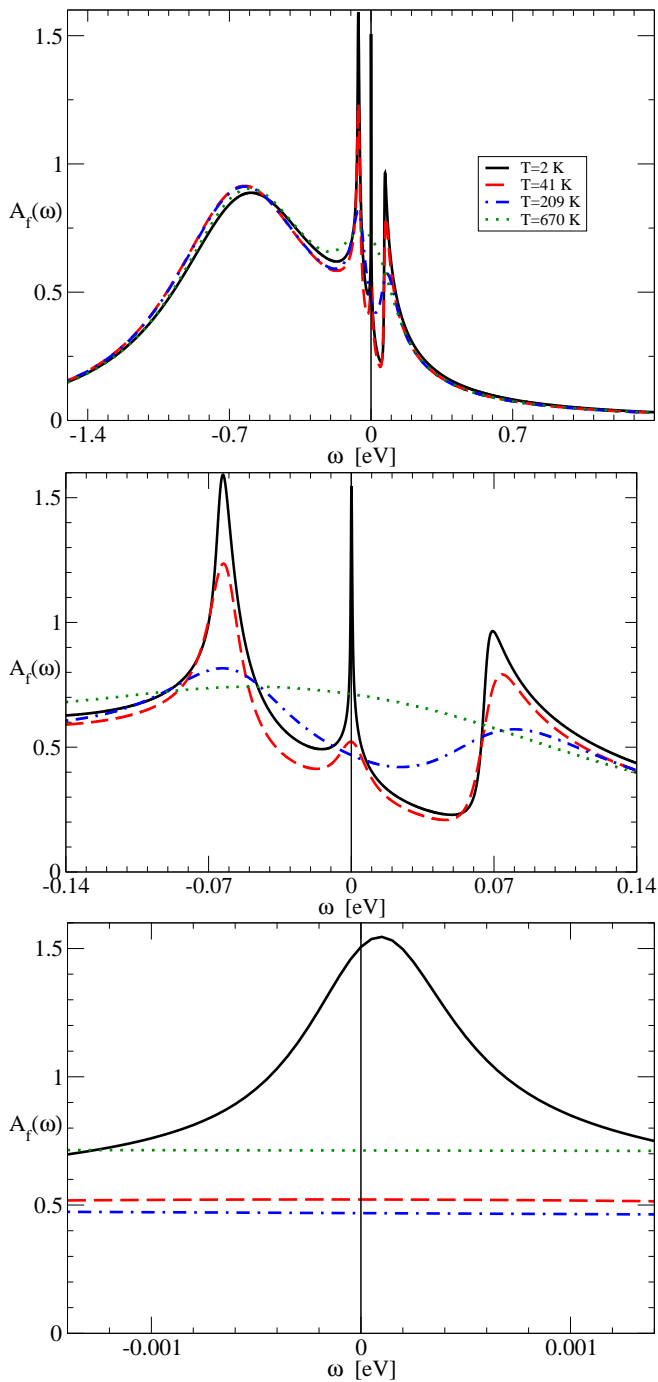

\center{
\includegraphics[width=1\columnwidth,clip]{./figures/A_0.06_new.eps}}                             
{\includegraphics[width=1\columnwidth,clip]{./figures/A_0.06_zoom_new.eps}}                
{\includegraphics[width=1\columnwidth,clip]{./figures/A_0.06_zoom_zoom_new.eps}}    
\caption{
Spectral function $A(\omega)$, calculated for the hybridization
strength $\Gamma$=0.06 eV and the CF splitting $\Delta$=0.07 eV,
plotted as a function of
frequency for several temperatures.
The solid, dashed, dashed-dotted and dotted curves correspond to $T=$2,
41, 209, and 670 K, respectively.
The charge-excitation peak is visible in the upper panel.
The  middle panel shows the evolution of the CF and Kondo resonances
with temperature.
For $T\leq \Delta$, the many-body resonance of half-width $\Delta$ is
centered well
below $\mu$. The F window has more states below than above $\mu$ and $S(T)<0$.
The lower panel shows the position of the Kondo resonance above $\mu$.
Its center defines $T_0=$1 K.  For $T\leq T_0$ the F window has more
states above than below $\mu$ and $S(T)>0$.
}
                                     \label{fig:spectrum_a}
\end{figure}

In this section we present the NCA results for $A(\omega)$, study
the low-energy spectral features in the vicinity of various fixed points,
discuss the changes induced by the crossovers, and
explain the behavior of $S(T)$ in terms of the redistribution of
spectral weight within the Fermi window (F window). Only the Ce case 
is considered and
it is assumed that pressure gives rise to an increase of the hybridization
width $\Gamma$.
The results obtained for $\Gamma <\Delta$ are shown in 
Fig.~\ref{fig:spectrum_a},
where $A(\omega)$ is plotted as a function of frequency for several
temperatures.
At high temperatures, $T\simeq T_\Delta$, the spectral function has a
broad charge-excitation peak somewhat above $E_f$ and
a narrower resonance of half-width $\Delta$, centered below $\mu$.
This low-energy resonance is a many-body effect due to the
hybridization of the conduction
states with the 4{\it f} states and is typical of the exchange 
scattering on the full multiplet.
In  this temperature range, the F window shows more spectral weight
below than above $\mu$ (see the middle panel in Fig. 
\ref{fig:spectrum_a}) and $S(T) < 0$.
The magnetic susceptibility\cite{BCW87}  is Curie-Weiss like,
with a very small Curie-Weiss temperature and a Curie constant
which is close to the free Ce ion value.
The maximum of $S(T)$ at about $T_S\simeq T_\Delta/2$ is here 
negative,  $S_{max}<0$,
but a slight increase of $\Gamma$ would make $S_{max}$  positive.
All these features are typical of the LM fixed point corresponding to 
a fully degenerate {\it f\/}-state.
At lower temperatures, $T < T_S$, the CF splits the many-body 
resonance into two peaks.
The larger one grows below $\mu$ and the smaller one above $\mu$
(see the middle panel in Fig. \ref{fig:spectrum_a}).
This asymmetry is enhanced as $\Gamma$ is reduced,
which is typical of the Anderson model with CF splittings;~\cite{BCW87}
the increase of the low-energy spectral weight below $\mu$ gives rise
to a large negative thermopower.
A further reduction of temperature leads, for $T \ll T_\Delta$, to a
rapid growth of an additional peak very close to $\mu$, such that
$A(\omega)$ acquires three pronounced low-energy peaks (see Fig. 
\ref{fig:spectrum_a}).
(The physical origin of these many-body resonances is explained in 
detail in Ref.~\cite{BCW87})
The peak centered at $\omega_0\ll\Delta$ is the Kondo resonance and
its appearance below $T\leq$ 40 K marks the onset of the LM regime,
which is due to scattering of conduction electrons on the lowest CF level.
The two CF peaks centered at about $\omega_0\pm\Delta$ are outside 
the F window,
and do not affect the low-temperature transport and  thermodynamics.
Once the Kondo peak peak  appears, the F window shows more spectral
weight above than below $\mu$ and $S(T)$ is positive, which is just 
the opposite
to what one finds for $T\geq T_\Delta$.
The center of the Kondo resonance saturates at low temperatures at the
energy $\omega_0>0$, which provides the NCA definition of the  Kondo
scale, $k_B T_0=\omega_0$.
In the symmetric Anderson model $T_0$ is related to the width
of the Kondo resonance but in the highly asymmetric case we are
dealing with here,  the current definition is more appropriate.
The comparison with numerical renormalization group (NRG)
calculations~\cite{Costi96}  shows that $\omega_0$ gives a reliable
estimate of the Kondo temperature even for a doubly degenerate
Anderson model, and we assume that the NCA definition of $T_0$
provides the correct Kondo scale of the CF-split single-impurity Anderson
model as well.
Because the Kondo resonance is asymmetric with respect to the $\omega=0$ line
and has more states above than below $\mu$, the reduction  of temperature
enhances $S(T)$ unltil it reaches, at $T_0$, the maximum value $S_0$.
A further temperature reduction leaves the top of the Kondo
resonance outside the F window, and the thermopower drops.
However,  most Ce and Yb system with very small $T_0$ have a phase transition
above $T_0$, and to discuss the normal-state properties of (a)-type systems
it is sufficient to consider the NCA solution for $T \geq T_0$.

\begin{figure}
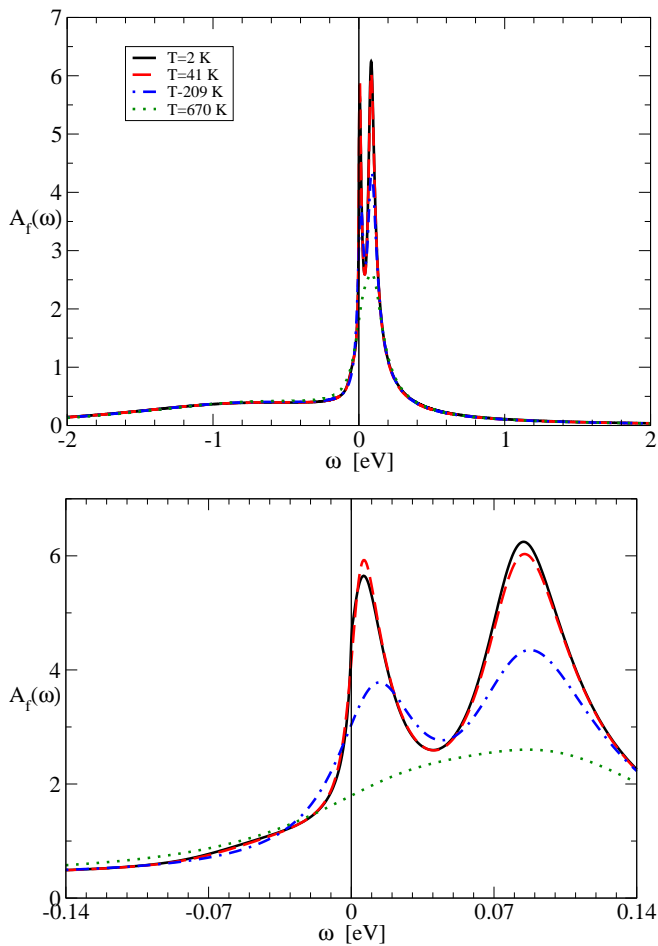

\center{\includegraphics[width=1\columnwidth,clip]{./figures/A_0.120_new.eps}}                      
\center{\includegraphics[width=1\columnwidth,clip]{./figures/A_0.120_zoom_new.eps}}          
\caption{
Spectral function $A(\omega)$, calculated for the hybridization
strength $\Gamma$=0.12 eV
and the CF splitting $\Delta$=0.07 eV, plotted as a function of
frequency for several temperatures.
The solid, dashed, dashed-dotted and dotted curves correspond to $T=$2,
41, 209, and 670 K, respectively.
The upper panel shows the overall features.
The two many body resonances are resolved but the lower CF peak and
the charge-excitation peak are absent.
The  lower panel shows the evolution of low-energy resonances with
temperature.
For $\Gamma > \Delta$, there is more spectral weight above than below
$\mu$ at all
temperatures and $S(T)$ is always positive.
}
                               \label{fig:spectrum_b}
\end{figure}

An increase of the coupling to $\Gamma > \Delta$ has a drastic effect 
on $A(\omega)$,
as illustrated in Fig.~\ref{fig:spectrum_b}, where $A(\omega)$ is plotted
for $\Gamma = 0.12$ eV.
The charge-excitation peak is transformed into a broad background
(see the upper panel in Fig.~\ref{fig:spectrum_b})
and the only prominent feature at $T\simeq T_\Delta$ is the low-energy
resonance of half-width $\Delta$ centered above $\mu$.
This low-energy resonance is due to the exchange scattering
of conduction electrons on the full CF multiplet,
which gives rise to the maximum of $S(T)$ in the LM regime.
The F window (see  lower panel in Fig.~\ref{fig:spectrum_b}) shows
more spectral weight above than below $\mu$, so that $S(T) > 0$.
The reduction of temperature below $T_S$ removes some spectral weight 
above $\mu$
and brings additional spectral weight below $\mu$, which reduces $S(T)$
and leads to a minimum.\cite{minimum}
A further reduction of temperature leads to the rapid growth of the
Kondo peak at $\omega_0$, and the CF peak at  $\omega_0+\Delta$,
but the negative CF peak does not develop. That is, an increase of pressure
removes the lower CF peak, and shifts the Kondo and the upper CF peak to higher
energies, without changing their separation $\Delta$.
The F window shows more spectral weight above than below $\mu$,
so that $S(T)$ is positive and grows as temperature is lowered.
The maximum $S_0$ is reached at $T_0$ when the Kondo resonance is 
fully developed.
The characteristic energy scale is defined again by the position of 
the Kondo peak,
$k_B T_0=\omega_0$, which can now be quite large.
For $T\leq  T_0$ the F window becomes narrower than the Kondo resonance and
$S(T)$ drops below $S_0$.
For $T\ll T_0$,  where the FL behavior is expected,\cite{BCW87}
the NCA gives $A(\omega)$ with an unphysical spike at $\mu$,
which makes $\rho_{mag}(T)$ and $S(T)$ much larger than the exact result.
However, once $T_0$ is obtained from the NCA calculations,
the low-temperature transport can be inferred from the universal power
laws which hold in the FL regime, as discussed in the previous section.

\begin{figure}
\center{\includegraphics[width=1\columnwidth,clip]{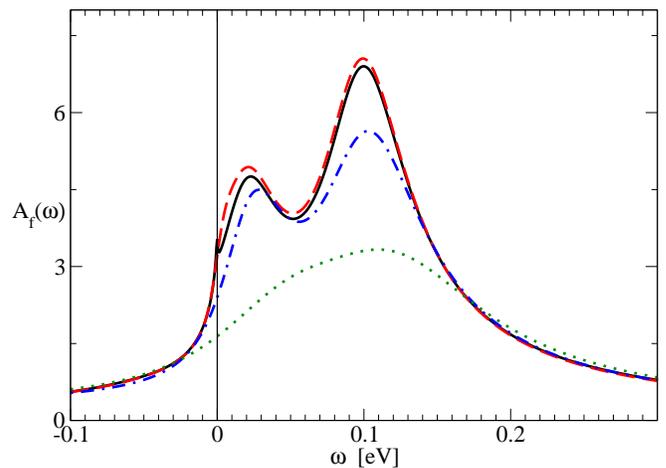}}    
\caption{Spectral function $A(\omega)$, calculated for the
hybridization strength
$\Gamma$=0.140 eV and the CF splitting $\Delta=0.07$ eV, plotted as a function
of  frequency for several temperatures. The solid, dashed,
dashed-dotted and dotted
curves correspond to $T$=2, 41, 209, and 670 K, respectively.
For $\Gamma > \Delta$, the Kondo resonance is reduced to a small hump
above the {\it f\/} level. }
                                \label{fig:spectrum_c}
\end{figure}

A further  increase of $\Gamma$ shifts the Kondo and the CF peaks to higher
energies, and changes  their relative spectral weight,
as  shown in Fig.~\ref{fig:spectrum_c}, where the low-frequency part 
of $A(\omega)$
is shown  for $\Gamma = 2\Delta$.
The Kondo scale is still defined by the center of the Kondo peak, 
even though it is now
reduced to a hump on the low-energy side of a large peak centered at 
$\omega_0+\Delta$.
The unphysical NCA spike at  $\omega=0$ can be seen at lowest temperatures.
The thermopower is positive at all temperatures and has only a
shoulder  below $T_S$. A quantitative comparison between
$T_0$, defined by the position of the Kondo resonance,  and the position
of the Kondo anomaly in $S(T)$ becomes difficult.

\begin{figure}
\center{\includegraphics[width=1\columnwidth,clip]{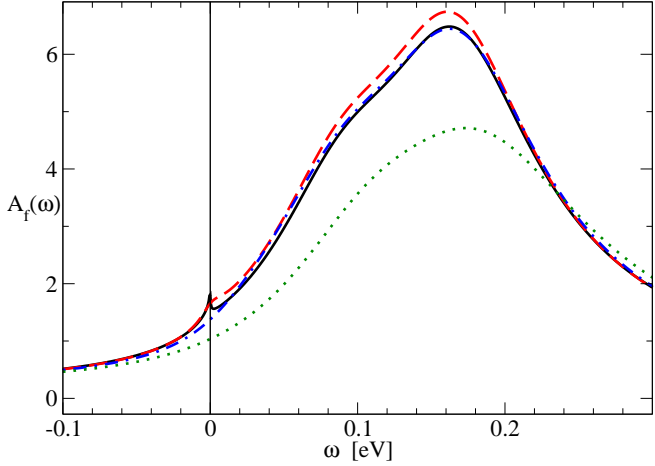}}     
\caption{Spectral function $A(\omega)$, calculated for the
hybridization strength
$\Gamma$=0.20 eV and the CF splitting $\Delta=0.07$ eV, plotted as a function
of  frequency for several temperatures. The solid, dashed,
dashed-dotted and dotted
curves correspond to T=2, 41, 209, and 670 K, respectively.
For $\Gamma\gg \Delta$, the Kondo resonance is absent.  }
                                \label{fig:spectrum_d}
\end{figure}

\begin{figure}
\center{\includegraphics[width=1 
\columnwidth,clip]{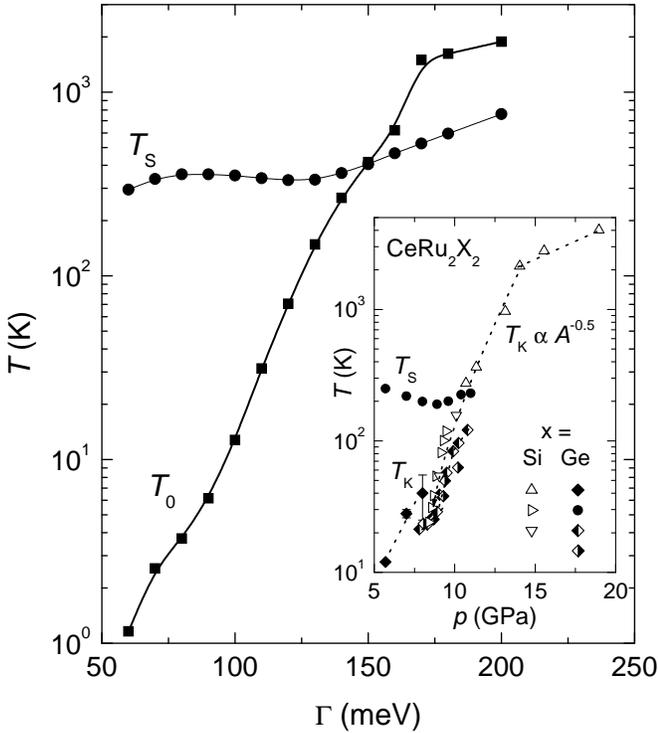}}                      
\caption{The temperature $T_S$, defined by the high-temperature
maximum of the thermopower, and the characteristic scale $T_0$,
defined by the peaks in the spectral function, are plotted
as functions of the hybridization strength $\Gamma$.  The three
regions, where $T_0$ changes its functional form, are clearly seen.
The inset reproduces the experimental data for
CeRu$_{2}$Ge$_2$ and CeCu$_{2}$Ge$_2$.\cite{Wilhelm04}
}
                      \label{T0-versus-gamma.eps}
\end{figure}

Finally, for $\Gamma > 2\Delta$, we find $A(\omega)$  with a single
broad peak centred at ${\tilde{E_f}}>0$, as shown in Fig.~\ref{fig:spectrum_d}.
The CF excitations are now absent,  which is typical for the Anderson model
in the vicinity of  the VF fixed point.
The relevant energy scale at low temperature is defined as $k_B T_0={\tilde{E_f}}$,
and shows an almost linear dependence on $ \Gamma$.
The thermopower is always positive and grows monotonically
from small values at low temperatures towards a  high-temperature
maximum at $T_S$.
The unphysical spike at $\omega=0$ appears at  higher temperatures
and is more pronounced than for small $\Gamma$, as illustrated in 
Fig.~\ref{fig:spectrum_d}.
The initial slope of $S(T)$ obtained from the NCA result for 
transport coefficients
is very much overestimated with respect to the FL result 
based on Eq.(\ref{resonant_SIAM}).
An increase of temperature above $T_S$ modifies the excitation spectrum
on an energy scale of the order of ${\tilde{E_f}}$ and reduces $S(T)$.
The values of ${\tilde{E_f}}$ and  $T_S$ do not seem to be related in
any simple way, and the temperature-induced breakdown of the local FL
is not indicated by any FL scale.
An increase of $\Gamma$ (pressure) changes $A(\omega)$ and enhances $T_S$
but does not lead to further enhancement of $S_{max}$.
The pressure dependence of ${\tilde{E_f}}$ and  $T_S$ looks different.
In other words, the description of  the VF regime requires more than 
one  energy scale.

The characteristic scales of the Anderson model obtained for
various values of $\Gamma$ provide the $(T,p)$ phase diagram
of the system plotted in Fig.~\ref{T0-versus-gamma.eps}.
The $T_0$ line is defined by the position of the Kondo peak and the 
$T_S$-line by  $S_{max}$.
At small $\Gamma$ (low pressure)  Fig.~\ref{T0-versus-gamma.eps}
indicates two crossovers. The one around  $T_0$ is between the FL and
the LM regime defined by the lowest CF level. The one around $T_S$ is 
the crossover
to the high-temperature LM regime, as defined by a full CF multiplet.
When these LM regimes are well separated,
the crossover between them is accompanied by a minimum and,
possibly, a sign-change of $S(T)$.
At intermediate pressures, the two LM regimes are too close for the
sign-change to occur,
and the crossover is indicated only by a shallow minimum or just a
shoulder of $S(T)$.
Here, the relationship between the low-temperature maximum of $S(T)$
and the center of the
Kondo peak can only be given as an order-of-magnitude estimate.
At very high pressures, the system is in the VF regime and the
crossover from a universal
low-temperature FL phase to a non-universal high-temperature phase
takes place at $T\simeq T_S$.
This crossover is not defined by the FL scale $ {\tilde{E_f}}$, which
is very large, but by a much smaller scale $T_S$.

\section{Conclusions  and summary
\label{Conclusions}  }
We have applied the single-impurity Anderson model to the investigation of the
temperature and pressure dependence of the thermopower of Ce and Yb 
intermetallics,
and found that  the crossovers between various fixed points explain 
the seemingly
complicated temperature dependence of $S(T)$.

The basic assumption of our approach is that for a given concentration of
rare-earth ions, and above some coherence temperature, the system is in
an "effective impurity" regime. In that "impurity limit", we treat 
the rare-earth
ions as independent scattering centers and solve the ensuing single-impurity
model  by the  NCA.
Our calculations impose the charge neutrality constraint on the local 
scattering
problem, which provides a minimal self-consistency condition for describing
the pressure effects. The excitation spectrum obtained in such a way is very
sensitive to the changes in the coupling constant $g=\Gamma/\pi|E_f|$,
and an increase of the hybridization or a shift of the {\it f} state, 
which modifies
the magnetic state of the {\it f} ion, has a huge effect on the 
spectral function.
Since the excitation spectrum is related to the transport coefficients by Kubo
formulas, the thermopower changes rapidly as a function of 
temperature, pressure
or doping. The shapes (a) to (d) assumed by  $S(T)$ in various regions
of the parameter space follow straightforwardly from the redistribution of the
single-particle spectral weight within the Fermi window.

The NCA solution of the single-impurity Anderson model breaks down 
for $T < T_0$.
However, the excitations above the ground state are determined by the 
general FL laws,
which can be used to find the initial slope of $S(T)$. Estimating the 
Kondo scale and
the number of {\it f} electrons by the NCA calculations, we find that 
an increase of pressure
reduces (enhances) the low-temperature value of $S(T)/T$ in Ce (Yb) compounds.
Combining the FL and the NCA results, we can obtain $S(T)$ in the full
temperature range at any pressure.

Our results explain the thermopower of most intermetallic  Ce compounds
mentioned in Section \ref{Introduction}.
The multi-peaked $S(T)$, which characterizes the type (a) and  (b) systems,
is obtained for Kondo ions with small $T_0$ and well-defined CF resonances.
A type (c) thermopower, which has weakly resolved peaks or just a broad hump,
is obtained for Kondo ions with large  $T_0$ and partially 
overlapping CF resonances.
A single-peaked $S(T)$ of the (d)-type is obtained for the VF ions which
do not show any CF splitting of the single-particle excitations.
If we assume that pressure increases $\Gamma$, and reduces $n_f$,
our results account for the changes of $T_0$, $S_0$, $S(T)/T$, and $S_{max}$,
observed in pressure experiments on Ce compounds. 
\cite{Jaccard85,jaccard.88,Wilhelm04,jaccard.92,jaccard.96,Link96b,Link96a}
The strong doping dependance\cite{sakurai.96} of $S(T)$ is explained
as a  chemical pressure effect, which changes the {\it f\/} occupation.
In our local model the "effective concentration" of {\it f} electrons is
determined self-consistently, and the thermopower of concentrated and
dilute systems cannot be related by simple scaling.  (Just like the 
high pressure data
cannot be described in terms of rescaled ambient pressure data.)
The relevant energy scales obtained for different values of $\Gamma$
agree with the $(T,p)$ phase diagram of CeRu$_{2}$Ge$_2$ and
CeCu$_{2}$Ge$_2$, inferred from the high-pressure transport and
thermodynamic data.\cite{Wilhelm04}

For Yb compounds, we argue that pressure or chemical pressure mainly
affect the position of the crystal field levels relative to the band center,
while leaving the hybridization width unchanged.
By shifting $E_f$, so as to increase the number of holes, we obtain the
thermopower of the type (a) to (d), in agreement with the experimental data.
The qualitative features seen in Yb intermetallics at various pressures or
chemical pressures are captured
as well,\cite{alami-yadri.99,nakamoto.99,casanova.90,trovarelli.99,andreica.99}
but different compounds require different initial parameters,
and a detailed analysis is yet to be done.

In summary, the normal-state properties of stoichiometric compounds
with Ce and Yb ions seem to be very well described by the local model
which takes into account spin and charge  fluctuations.
The classification of the thermopower data follows straightforwardly
from the fixed point analysis of the single-impurity Anderson model
with the CF splitting.  A rich variety of shapes assumed by $S(T)$ at
various pressures or doping dramatically illustrates the effect of the
local environment on the response of the single rare earth ion.
The nature of the ground state, and the fact that the ground state of 
many Ce and
Yb systems can be changed with pressure or doping, do not seem to affect the
thermopower  in the normal state.  We take this as an indication that 
the normal
state properties of Ce and Yb compounds are dominated by the local dynamics.

The main problem with our poor man's treatment of ordered compounds
is that it neglects  the coherent scattering which sets in at low enough
temperatures, and reduces the thermopower below the values predicted
by the local Fermi liquid theory.
We can argue that these effects do not change the qualitative features of the
thermopower but the proper answer cannot be obtained without solving
the lattice model.

{\bf Acknowledgements}\\
We acknowledge useful comments and suggestions from
B. Coqblin, J. Freericks, C. Geibel, A. Hewson, B. Horvati\'c, and H. Wilhelm.
This work has been supported by the
Ministry of Science of Croatia (CRO-US joint projects, grant number 1/2003,
by the National Science Foundation under grant number DMR-0210717,
and the Swiss National Science Foundation under grant number 7KRPJ065554-01/1.
\vspace*{5mm}



\begin{thebibliography}{99}
\bibitem{n_f=0}
For $\Gamma\ll \Delta$, such that $n_f\to 1$, the NCA equations break
  down even  before  $T=T_0$ is reached.
\bibitem{quartet}
The results for two near-by excited doublets are the same.
\bibitem{minimum}
For not too large values of  $\Gamma$, the thermopower at the minimum 
is negative
(see $\Gamma$=90 meV curve in  Fig.~\ref{fig:theo_tep_low}) but if 
$\Gamma$ continues
to increase the minimum of $S(T)$ becomes more shallow and does not extend
to negative values  (see curves in  Fig.~\ref{fig:theo_tep_low} 
obtained for $\Gamma$=100-120 meV).
\end{thebibliography}
\end{document}